\documentclass[prb,preprint]{revtex4-1} 
\usepackage{graphicx} 
\usepackage{epsf}
\newcommand{\be}{\begin{equation}}
\newcommand{\ee}{\end{equation}}
\newcommand{\ba}{\begin{eqnarray}}
\newcommand{\ea}{\end{eqnarray}}
\newcommand{\nn}{\nonumber}
\newcommand{\la}{\label} 
\newcommand{\e}{{\rm e}}
\newcommand{\h}{{\cal H}}
\newcommand{\w}{\omega} 
\newcommand{\m}{{\bf M}} 
\newcommand{\dt}{\epsilon} 
\newcommand{\ep}{\epsilon} 
\newcommand{\bx}{{\bf X}}
\newcommand{\bxp}{{\bf X}^\prime}
\newcommand{\x}{{\bf x}}
\newcommand{\xp}{{\bf x}^\prime}
\newcommand{\R}{{\bf X}}
\newcommand{\br}{{\bf x}}
\newcommand{\rr}{{\bf r}}

\begin{document}

\title{High-order Path Integral Monte Carlo methods for solving quantum dot problems }

\author{Siu A. Chin}
\affiliation{Department of Physics and Astronomy, Texas A\&M University,
College Station, TX 77843, USA}

\email{chin@physics.tamu.edu} %

\begin{abstract}
The conventional second-order 
Path Integral Monte Carlo method is plagued with the sign problem in solving many-fermion systems. This is
due to the large number of anti-symmetric free fermion propagators that are needed to 
extract the ground state wave function at large imaginary time.  In this work, we show 
that optimized fourth-order Path Integral Monte Carlo methods, which use no more than 
5 free-fermion propagators, can yield accurate quantum dot energies for up to 
20 polarized electrons with the use of the Hamiltonian energy estimator.  

\end{abstract}

\maketitle 


The Path Integral Monte Carlo (PIMC) method remains problemic for
solving many-fermion systems due to the ``sign" problem.
Since the anti-symmetric free-fermion propagator (FFP) is not positive-definite, only
its magnitude can be sampled by the Monte Carlo method and observables must then be weighted
by the overall sign of {\it all} FFP in the path-integral.
While a suggested remedy has been proposed\cite{mak98,egg99,egg00}, the most practical solution
in traditional PIMC is to simply 
side-step the sign-problem by invoking some ``fixed-node" or 
``restricted-path" approximations\cite{cep95}. Here, we argue that: 1) The sign-problem
is {\it not} intrinsic to solving a fermion problem; 2) it is only a consequence of 
a {\it poor} approximation to the exact propagator; and 3) it can be automatically {\it minimized} 
by using better, higher-order approximate propagators. 
By following up on the last point, this work
shows that by using optimize fourth-order propagators, accurate results can be obtained
for up to 20 spin-polarized electrons in a 2D circular, parabolic quantum dot.  

\begin{figure}
	\vspace{0.5truein}
	\centerline{\includegraphics[width=0.9\linewidth]{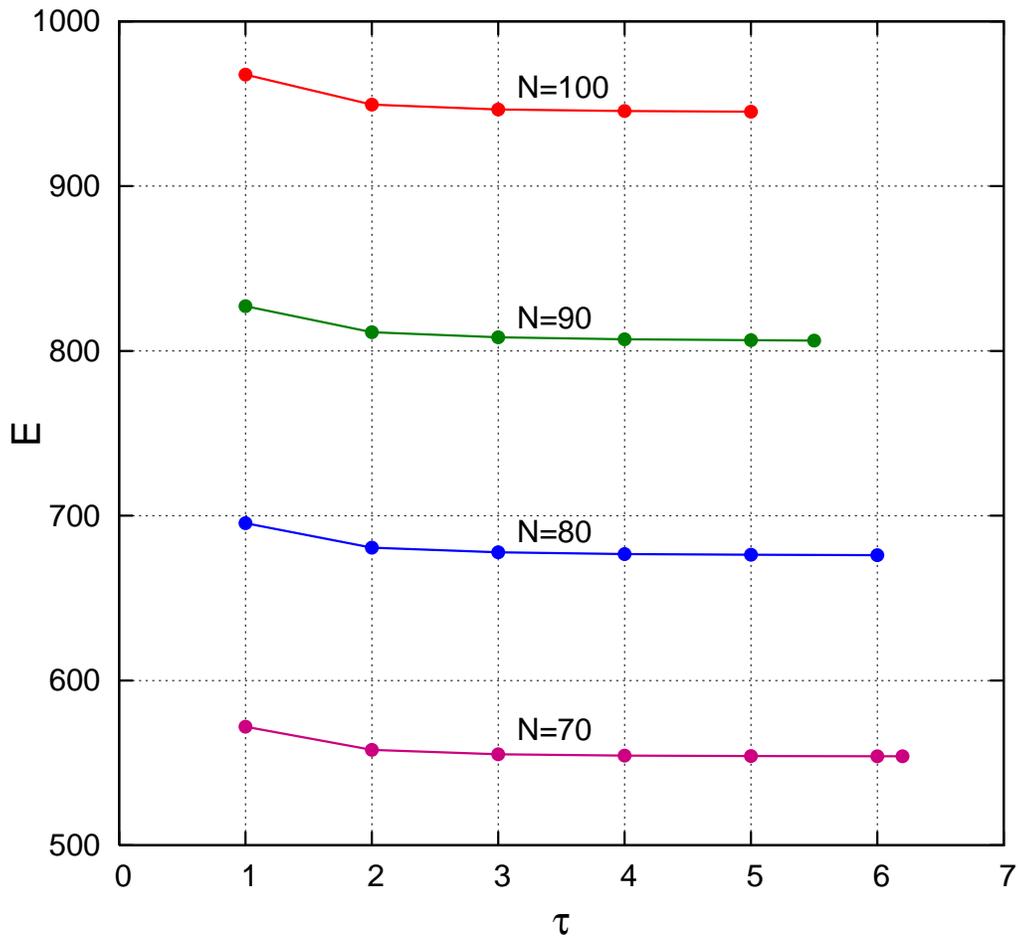}}
	\vspace{0.5truein}
\caption{A two-propagator calculation using the exact, HO 
fermion propagator (\ref{exho}) in 2D. $E$ is the energy in units of $\hbar\omega$ and $\tau=\beta\omega$.
Symbol sizes are larger than statistical error bars. For 70 to 100 fermions, the exact energies 
are 554, 676, 806 and 945. The results obtained here at the largest value of $\tau$ are
554.044(5) 676.045(9), 806.312(6) and 945.22(2) respectively.
The residual errors are not statistical, but are due to software precision limiting  
the maximum value of $\tau$ that can be reached. See text for details. 
 \label{ho}}
\end{figure}

First, let's dispel the notion that the sign-problem is intrinsic to solving fermion 
problems in PIMC. If the exact propagator of the system 
$G(\bx,\bxp;\tau)=\langle \bx| \e^{-\tau H}|\bxp\rangle$ is known, 
then one can compute the energy from 
\be
E=\frac{\int d\bx H G(\bx,\bx;\tau)}{\int d\bx G(\{\bx,\bx;\tau)}
\ee
without any sign-problem, since $G(\bx,\bx;\tau)>0$. To drive home this point with a
non-trivial example, we show in Fig.\ref{ho}, the ground state energies of up to 100 {\it non-interacting}
fermions in a 2D harmonic oscillator (HO). The exact many-fermion propagator 
is obtained by generalizing the HO propagator given in Ref.\onlinecite{chin05} to
\be
G(\bx,\bx^\prime;\tau)
=
(2\pi\tau C_T)^{-DN/2}\e^{-\tau C_V\frac12 \w^2\sum_{i=1}^N(\x_i^2+\x_i^{\prime 2})}
\det [\e^{-\frac1{2\tau C_T}(\x_i-\x_j^\prime)^2}],
\la{exho}
\ee
where $\bx=\{\x_i\}$,
$C_V=(\cosh(\w\tau)-1)/(\sinh(\w\tau)\w\tau)$,
$C_T=\sinh(\w\tau)/(\w\tau)$, $D$ is the dimension of $\x_i$ and $N$ is the
number of particles. (A proof of this will be given elsewhere, but it is easy to verify 
that it is exact by a direct calculation, as done here.) Fig.\ref{ho} demonstrated that this fermion problem can 
be solved by the Monte Carlo method without any sign-problem and the energy obtained is precisely 
that of filling the 2D HO energy levels up to the number fermions. (For $N\le 30$, 
a single $G(\bx,\bx;\tau)$ is adequate. For larger $N$, Fig.\ref{ho} uses two 
propagators $G(\bxp,\bx;\tau/2)G(\bx,\bxp;\tau/2)>0$. This is because at $\tau=6$, $\tau C_T\approx 200$, 
all entries of the matrix in (\ref{exho}) are nearly one, and the subtraction needed to 
compute the determinant is beyond Fortran's double-precision. With two propagators, the needed 
value of $\tau$ is halved.)

If one tries to approximate this, or any other exact propagator, by a product 
of primitive action (PA) second-order propagator having one FFP, then one is again
confronted by the sign-problem. Since the PA propagator is only accurate at very small time steps, 
conventionally hundreds of them are needed to extract the ground state 
at large imaginary times. When one has hundreds of FFP, then whether one is sampling the 
permuations\cite{cep95} or directly evaluating the anti-symmetric free-propagator\cite{ti}, 
the average of the sign is close to zero and the sign-problem is intractable. This
is in spite of the fact that the same problem has just been solve in Fig.\ref{ho} without
any sign-problem. From this perspective, {\it the sign-problem is purely a consequence of a poor 
approximation} to the exact propagator.
To the extent that one can better approximate the exact propagator, 
one can automatically lessen the sign problem. 
This also means that there is no imperative to ``solve"
the sign-problem. (Why insist on solving a problem using a poor approximation?)
The sign-problem is to be avoided by approximating the exact propagator more accurately
using the fewest FFP. If the number of FFP can be kept to less than 10, 
then one can extract ground state properties {\it before} the onset of the sign problem. 
This is the key idea of this work.
  
This idea has not been contemplated before because traditional PIMC has
been formulated mostly in terms of the low-order, inaccurate PA propagator.
It is therefore always deemed impossible to reach the ground state with so few FFP. 
However, in the bosonic case of liquid helium, we have shown\cite{cif03} 
that even a single use of a {\it fourth-order} propagator can reach closer to the ground state than many
elaborate variational schemes. Recently, the ground state energy of liquid Helium has been computed in PIMC 
by Sakkos, Casulleras and Boronat\cite{sak09}, using a fourth-order propagator
that can be ``fine-tuned" to converge at the sixth-order for the energy. 
Their work reduces, by an order-of-magnitude, the number of propagators (or beads) 
needed in bosonic PIMC. Subsequently, R. Rota {\it et al.}\cite{rota10} 
showed that a Ground-State PIMC (GSPIMC) calculation only needed about ten propagators to achieve the same
objective! Similar results have also been obtained for the same system with GSPIMC by Zillich, Mayrhofer 
and Chin\cite{zill10}, using sixth and eighth-order extrapolated propagators. The tremendous sucess 
of these results have inspired this work to apply higher-order PIMC methods 
to fermions in quantum dots. Below we will derive fourth-order algorithms for solving
a general interacting fermion problem, where the exact HO propagator (\ref{exho}) 
is not used. (Since the electron repulsion expands the size of the quantum dot,
the exact HO propagator is also not effective at strong couplings.)

Aside from using higher-order propagators, the above calculations\cite{cif03,rota10,zill10} 
also computed the energy directly from the Hamiltonian.
By the Golden-Thompson inequality\cite{gold65,thom65}, the {\it thermodynamic} estimator
used in conventional PIMC converges to the ground state energy only from below. 
In the few-propagators case, the thermodynamic estimator is so far from
convergence that it is totally useless. The use of the {\it virial} estimator is risky
for quantum dots, since it may dip below the exact ground state energy\cite{egg99}. In this work,
we follow the success of the GSPIMC method in also using the {\it Hamiltonian} estimator in PIMC. 
This estimator is known\cite{whit10} in Bosonic PIMC, but we generalize it here to include 
the anti-symmetric, determinant propagator. There are three advantages in using
the Hamiltonian estimator in PIMC. First, it gives a variational upper-bound to the ground state 
energy, as in GSPIMC. Second, its result can be double-checked by use of a variant, 
the Clark-Westhaus\cite{jc79} (CW) form of the kinetic energy. Third, in contrast to GSPIMC, 
no trial ground-state wave function is needed.  

The Hamiltonian for $N$ electrons in a 2D harmonic dot is
\be
H=\sum_{i=1}^N(-\frac{\hbar^2}{2 m^*}\nabla^2_i+\frac12 m^*\omega^2\rr_i^2)+\sum_{i<j}^N\frac{e^2}{\kappa|\rr_i-\rr_j|}.
\ee
By expressing $\rr_i=\ell_0\, \x_i$, where $\ell_0=\sqrt{\hbar/(m^*\omega)}$ is the harmonic length,
one obtains the dimensionless Hamiltonian $\h$ in terms of dimensionless vectors $\x_i$
\be
\h\equiv\frac{H}{\hbar\omega}=
\sum_{i=1}^N(-\frac12 \nabla^2_i+\frac{1}{2}\x_i^2)
+\sum_{i<j}^N\frac{\lambda}{|\x_i-\x_j|}
\la{dham}
\ee
with effective coupling strength
$
\lambda=\ell_0{m^*e^2}/(\hbar^2\kappa)=l_0a_B^{-1},
$
where $a_B$ is the effective Bohr radius. The Hamiltonian ${\cal H}$ is the
quantum dot's energy in units of $\hbar\w$.

The corresponding imaginary time propagator is
\be
G(\tau)=\e^{-\beta H/\hbar}=\e^{-\tau\h}=\e^{-\tau(T+V)},
\la{fpg}
\ee
where $\tau=\beta\omega$ is the dimensionless imaginary time, and
$T$ and $V$ are the kinetic and potential operators of $\h$.
The PA propagator approximates $G(\ep)$ 
at small time $\ep=\tau/n$ as
\be
  G_2(\ep) =\e^{-\ep V/2}\e^{-\ep T}\e^{-\ep V/2} +O(\ep^3),
\ee
with coordinate representation
\ba
G_2(\bx,\bxp;\ep)&=&\langle\bx|G_2(\ep)|\bxp\rangle\nn\\ 
&=&\e^{-\ep V(\bx)/2}G_0(\bx,\bxp;\ep)\e^{-\ep V(\bxp)/2},
\ea
where $\bx=\{\x_i\}$ is the position vector of all $N$ fermions, and 
\be
V(\bx)=\sum_{i=1}^N\frac{1}{2}\x_i^2
+\sum_{i<j}^N\frac{\lambda}{|\x_i-\x_j|}.
\ee
The anti-symmetric FFP $G_0(\bx,\bxp;\ep)$ is given by
\be
 G_0(\bx,\bxp;\ep)=\langle\bx|\e^{-\ep T}|\bxp\rangle
 =(2\pi\ep)^{-ND/2}\det\m ,
\ee
where $N$ is the number of fermions (electrons of the same spin),
$D$ is the dimension of the system, and $\m$ is the $N\times N$ anti-symmetric 
diffusion matrix
\be
M_{ij}(\bx,\bxp)=\exp\left[-\frac1{2\ep}(\x_i-\xp_j)^2\right].
\la{det}
\ee
For computing the energy, it is convenient to write 
\ba
G_0(\bx,\bxp;\ep)&=&\e^{-u_0(\bx,\bxp;\ep)},\nn\\
 u_0(\bx,\bxp;\ep)&=&\frac{ND}2 \ln(\ep)-\ln(\det\m).
\ea

In PIMC, the energy is calculated from ($\bx_k$ is denoted simply by $k$)
\be
E=\frac{\int d1...dn G_2(1,2;\ep)\h G_2(2,3;\ep)\cdots G_2(n,1;\ep)}
{\int d1...dn G_2(1,2;\ep)G_2(2,3;\ep)\cdots G_2(n,1;\ep)}
\ee 
and averaged over all $n$ places where $\h$ can be inserted between propagators. 
Since the FFP is not positive-definite, the above integral is sampled as
\be
E=\frac{\int d1...dn\ {\rm sgn}  E_{\rm H}(k,k+1) P(1,2,\cdots n;\ep)}
{\int d1...dn\ {\rm sgn} P(1,2,\cdots n;\ep)}
\ee 
with the probability distribution function taken to be 
\be
P(1,2,\cdots n;\ep)= |G_2(1,2;\ep)G_2(2,3;\ep)\cdots G_2(n,1;\ep)|,
\ee
and where ${\rm sgn}=\pm 1$ is the overall sign of the product of $G_2$'s. The Hamiltonian energy estimator
is given by 
\ba
&&E_{\rm H}(\bx,\bx^\prime;\ep)=\frac{\h G_2(\bx,\bx^\prime;\ep)}{G_2(\bx,\bx^\prime;\ep)}\nn\\
&&\qquad\qquad =\frac{ \sum_{i=1}^N(-\frac12\nabla^2_i) G_2(\bx,\bx^\prime;\ep)}{G_2(\bx,\bx^\prime;\ep)}+V(\bx)\ \ \ 
\ea
The alternative CW form of the kinetic energy is to let one of the gradient operator
acts to the {\it left}, giving
\ba
&&E_{\rm CW}(\bx^*,\bx,\bxp;\ep)\nn\\
&=& \frac{\sum_{i=1}^N \frac12 G_2(\bx^*,\bx;\ep)
{\buildrel \leftarrow\over\nabla_i}\cdot \nabla_i G_2(\bx,\bxp;\ep)}
{G_2(\bx^*,\bx;\ep)G_2(\bx,\bxp;\ep)}+V(\bx).\ \ \ \
\ea
The exact (and the free) propagator satisfies the equation
\be
-\frac{\partial}{\partial \ep}G(\bx,\bxp;\ep)=\h G(\bx,\bxp;\ep).
\ee
This equality no longer holds when $G(\bx,\bxp;\ep)$ is replaced by an approximation, 
such as $G_2(\bx,\bxp;\ep)$.
In this case, when both sides of the equation are divided by
$G_2(\bx,\bxp;\ep)$, the RHS gives the ``Hamiltonian" estimator as stated above.
The LHS then gives the ``Thermodynamics" energy estimator
\ba
&&E_{\rm TH}(\bx,\bx^\prime;\ep)=\frac{-\partial_\ep G_2(\bx,\bx^\prime;\ep)}{G_2(\bx,\bx^\prime;\ep)}\nn\\
&&\qquad\qquad =\frac{\partial}{\partial \ep}u_0(\bx,\bxp;\ep)+\frac12[V(\bx)+V(\bxp)].\ \ \ 
\ea
By the repeated use of the identity
\be
\frac{\partial}{\partial \alpha}\ln({\det\bf M})\nn\\
= {\rm Tr}\Bigl[{\bf M}^{-1}\frac{\partial {\bf M}}{\partial \alpha}\Bigr],
\ee
all three estimators can be computed without difficulties:
\ba
E_{\rm TH}(\R,\R';\dt)
&=&\frac{ND}{2\ep}-\frac1{2\dt^2}\sum_{i=1}^N(\x_i^2+{\x_i^\prime}^2-2\x_i\cdot\widetilde\x_i^\prime)\nn\\
&&\qquad\qquad +\frac12[V(\bx)+V(\bxp)],\la{eth1}
\ea
\ba
E_{\rm H}(\bx,\bx^\prime;\ep)
&=&\frac12\sum_{i=1}^N\nabla_i^2\Bigl[u_0+\frac\ep{2}V\Bigl]\nn\\
&&-\frac12\sum_{i=1}^N\Bigl[\nabla_i(u_0+\frac\ep{2}V)\Bigr]^2+V(\bx),\ \ \ 
\la{h1}
\ea
\ba
&&E_{\rm CW}(\bx^*,\bx,\bx^\prime;\ep)=\nn\\
&&\frac12\sum_{i=1}^N\nabla_i\Bigl[u_0(\bx,\bx^*;\ep)+\frac\ep{2}V\Bigr]
\cdot\nabla_i\Bigl[u_0(\bx,\bxp;\ep)+\frac\ep{2}V\Bigr]\nn\\
&&\qquad\qquad\qquad\qquad\qquad\qquad+V(\bx),
\la{cw1}
\ea
where
\ba
\nabla_iu_0(\bx,\bxp;\ep)
&=&\frac1{\dt}(\br_i-\widetilde\br'_i),
\la{gradu}\nn\\
\nabla_i^2u_0(\bx,\bxp;\ep)
&=&\frac{D}{\dt}-\frac1{\dt^2}\Bigl({\br'_i}^2-
\widetilde{\br}_i^{\prime 2}\Bigr).
\la{lap}
\ea
and where $\widetilde\br'_i$ is defined by 
\be
\widetilde\br'_i\equiv\sum_{k=1}^N\br'_k M_{ik}(\bx,\bxp)M^{-1}_{ki}(\bx,\bxp).
\ee
Thus in all three energy estimates, the calculation of $\m^{-1}$ is required. 
In the free propagator case, one has indeed $E_{\rm TH}=E_{\rm H}$. The
CW estimator will generally have greater variance than the Hamiltonian estimator.

\begin{figure}
	\centerline{\includegraphics[width=0.95\linewidth]{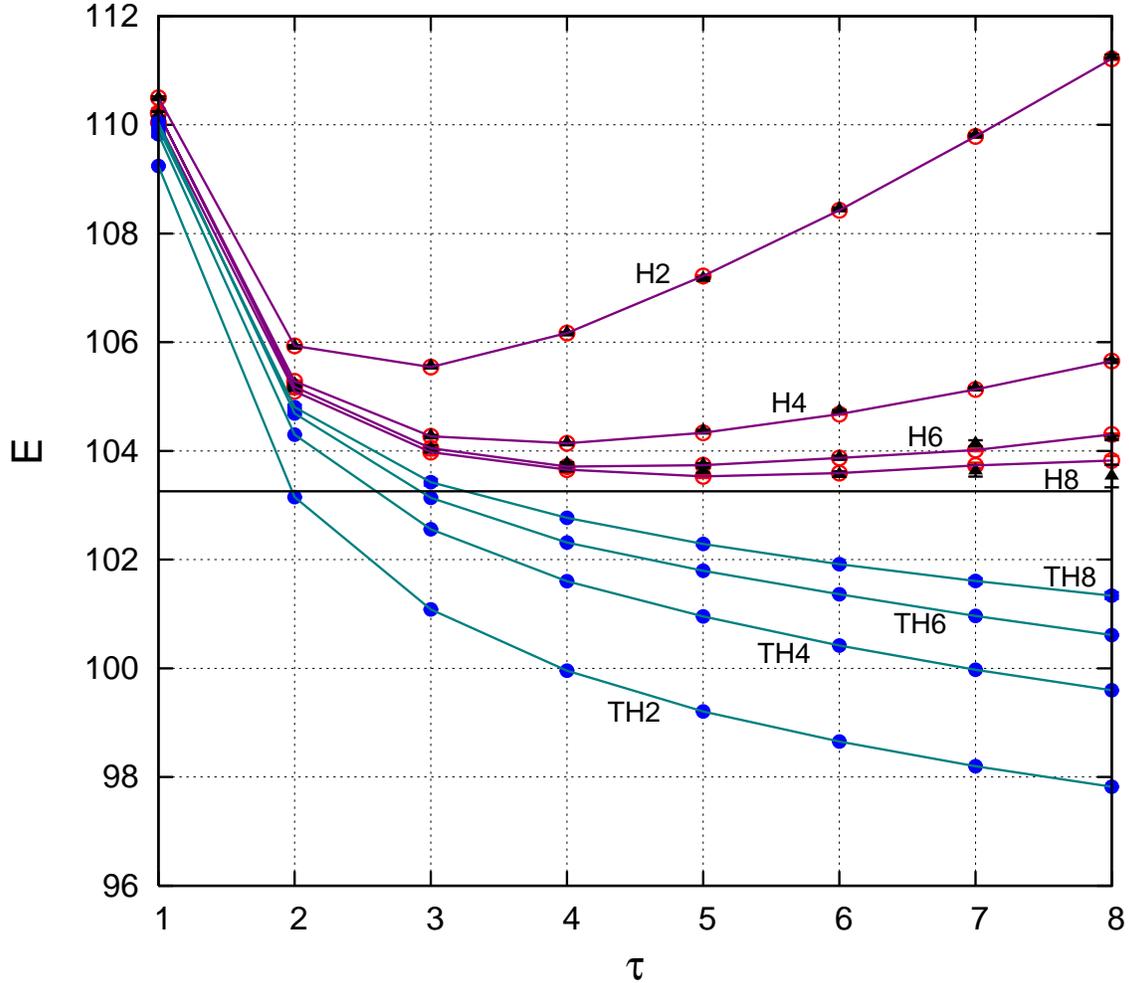}}
\caption{The convergences of the Hamiltonian energy (H) vs the Thermodynamics energy (TH)
in a second-order PA PIMC calculation for $N=8$ polarized
electrons at $\lambda=8$. The number is the number of PA propagators used
in that calculation.  The black-triangle and red-circles are the CW and Hamiltonian
energies respectively. The errorbars are computed from 60-100 block
averages of 5$\times 10^4$ configurations, and are mostly smaller than the plotting symbols. 
The black line is the PIMC result of Egger {\it et al.}\cite{egg99}. 
\label{pa}}
\end{figure}
 
After a set of $M$ configureations $\{\bx_i^{(m)}\}$ has been generated according to 
$P(\bx_1,\bx_2,\cdots \bx_n;\ep)$,
the energy can be computed by using the above three estimators as
\be
E=\frac{\sum_{m=1}^M {\rm sgn}_k [\frac1{n}\sum_{k=1}^n E_{\rm H,CW,TH}(\bx_k^{(m)},\bx_{k+1}^{(m)};\ep)]}
{\sum_{k=1}^M {\rm sgn}_k }.
\ee
For $N=8$ spin-polarized electrons at the strong-coupling limit of $\lambda=8$, the 
the energy of these three estimators are compared in Fig.\ref{pa}. 
This is the largest quantum dot at the strongest coupling considered in Egger {\it et al.}'s 
PIMC calculation\cite{egg99}
and in Rontani {\it et al.}'s configuration-interaction study\cite{ron06}. 
The thermodynamics estimator showed no convergence for up to 8 PA propagators, 
whereas the Hamiltonian and CW estimators are in excellent agreement in providing upper-bounds 
to the ground state energy from 2 to 8 propagators. The sign problem is completely under
control in these calculations. The Hamiltonian energy minimum in the
8 propagators case is already close to the result of Egger {\it et al.}\cite{egg99}.
This strongly suggests that improving the propagator beyond second-order can circumvent 
the sign problem in these quantum dot calculations.  

\begin{figure}
	\centerline{\includegraphics[width=0.95\linewidth]{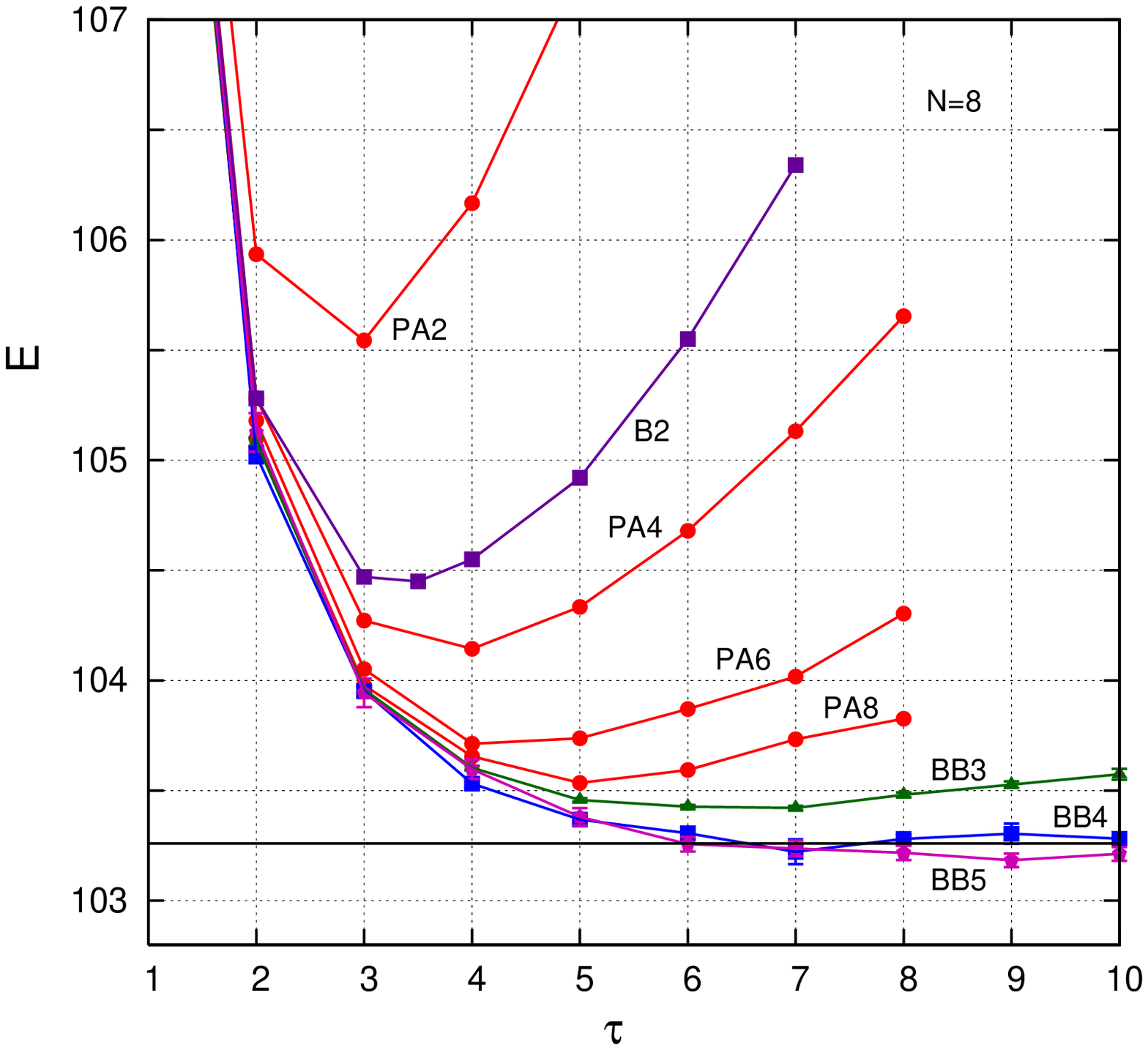}}
\caption{Comparing the Hamiltonian energies from optimized fourth-order 
propagators vs the PA Hamiltonian energies from Fig.\ref{pa}, for the same
$N=8$ quantum dot. The optimized propagator results are labeled as 
``Best Bead" (BB), with 3-5 free-fermion propagators. B2 has no
free-parameter to optimize the energy.
\label{compare}}
\end{figure}

The short-time propagator can be approximated
to any order by a product decomposition,
\be
{\rm e}^{-\ep (T+V )}=\prod_{i=1}^n
{\rm e}^{-t_i\ep T}{\rm e}^{-v_i\ep V},
\label{arb}
\ee
with a suitable set of coefficients $\{t_i, v_i\}$.
However, as first shown by Sheng\cite{sheng},
Suzuki\cite{suzukinogo}, Goldman-Kaper\cite{goldman}, and
more recently in a constructive proof by
Chin\cite{chin063}, beyond second order, any factorization of the form (\ref{arb})
{\it must} contain some negative
coefficients in the set $\{t_i, v_i\}$ and cannot be used in PIMC.
This is because if $t_i$ were negative, then replacing $\ep\rightarrow t_i\ep$ 
in the free fermion propagator (\ref{det}) would result in an unbounded
function that cannot be normalized as a probability. This 
simply reflects the fact that diffusion is a time-irreversible process.
To have {\it forward} fourth-order schemes, with all positive coefficients,
one must include the gradient potential
\be
[V,[T,V]]=\sum_{i=1}^N |\nabla_i V|^2,
\ee 
in the decomposing process\cite{suzfour,chin97}.
These fourth-order schemes have been used successfully in bosonic
DMC and PIMC simulations\cite{fchinl,jang,sak09,rota10}. Here, we will use a more
extended family of these forward fourth-order propagators, with arbitrary
numbers of free-fermion $T$ operators as described in Refs.\onlinecite{chin063,chin062}.
\begin{figure}
	\centerline{\includegraphics[width=0.95\linewidth]{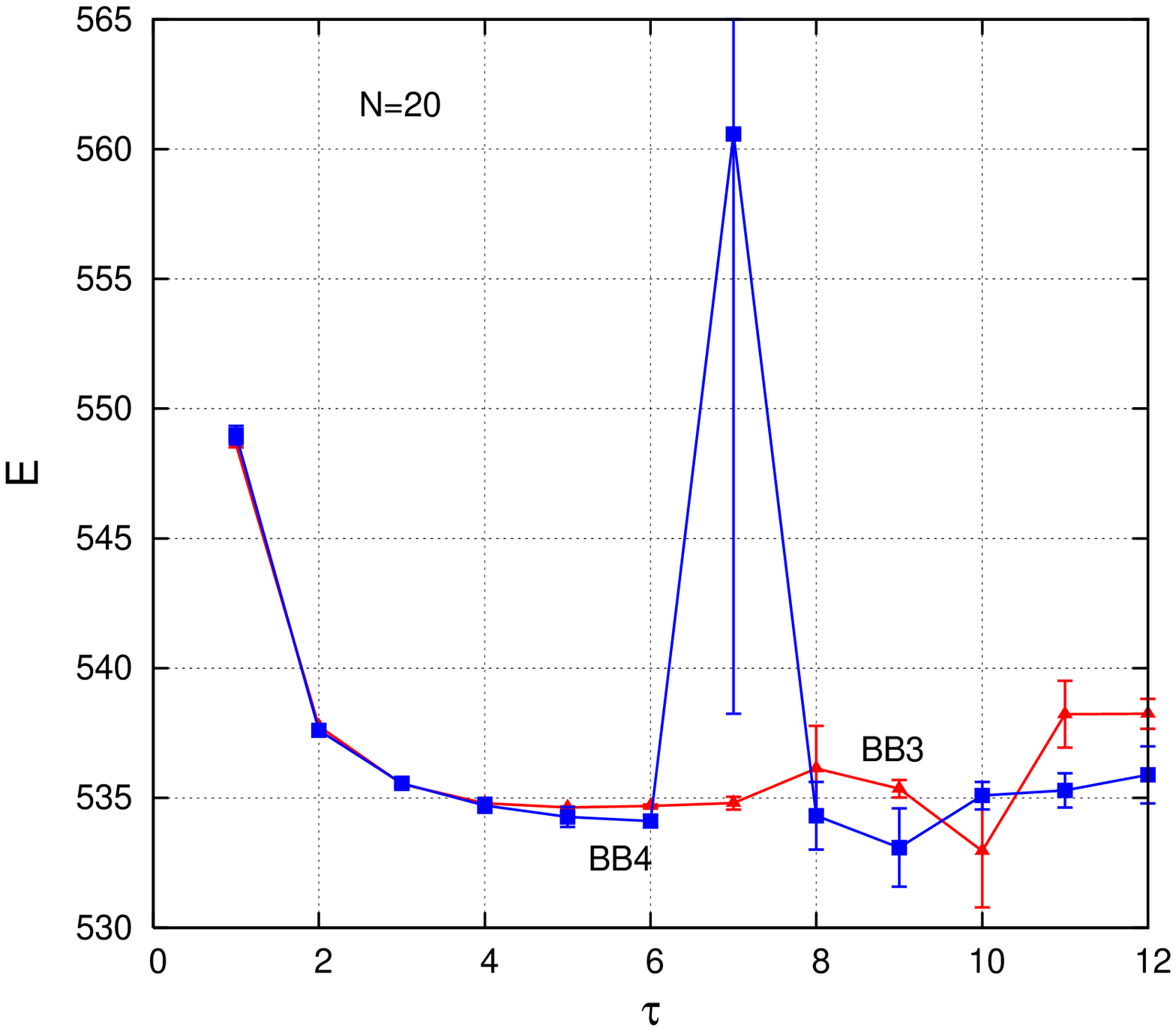}}
\caption{ Convergence of the Hamiltionian energy for $N=20$ polarized electrons using
the optimized, fourth-order 3- and 4-bead propagators. Errorbars are computed from
200-300 block-average of $5\times 10^4$ configurations of all 20 electrons.
\label{n20opt}}
\end{figure}

In order to compare with the PA algorithm,
we will characterize these algorithms by their number of $T$ operators, or beads.   
We will consider approximations of the form (\ref{arb}) 
with $t_1=0$ and with left-right symmetric coefficients $v_1=v_{N}$, $t_2=t_{N}$, etc.,
\ba
{\cal T}_{(N-1)B}^{(4)}(\epsilon)
&=&\e^{v_1\epsilon V}
  \e^{t_2\epsilon T}\e^{v_2\epsilon V}
  \cdots \e^{t_{N}\epsilon T}\e^{v_{N}\epsilon V}.
  \la{tb}
\ea
This will be fourth-order if one chooses $\{t_i\}>0$ with
$\sum_{i=1}^{N}t_i=1$, fixes $\{v_i\}$ by
\be
v_1=v_N=\frac12 + \lambda_2(1-t_2),\quad v_i=-\lambda_2(t_i+t_{i+1}),
\la{vcof}
\ee
where 
$\lambda_2=-\phi^{-1}/2$, $\ \phi=1-\sum_{i=1}^{N}t_i^3$, and divide 
the required gradient potential term
\be
\frac1{24}(\frac1{\phi}-1)\epsilon^3[V,[T,V]]
\la{vtv}
\ee
left-right symmetrically among all the $v_i\ep V$ terms in (\ref{tb}).
In order to compute the Hamiltonian energy as simply as in the PA case
(with only minor changes from $\ep\rightarrow t_i\ep$ and $\ep\rightarrow v_i\ep$),
the gradient potential term must {\it not} be distributed to the $v_1$ and $v_N$ 
potential terms. Because of this constraint, there is no free parameter in the
2-bead calculation (algorithm 4A of Ref.\onlinecite{chin97}) that can be used
to optimized the energy. Similarily, the Takahashi and Imada fourth-order trace propagator\cite{ti} 
cannot be used, becuase it is difficult to compute the Hamiltonian estimator with the gradient potential.
The freedom to choose $\{t_i\}$
and to distribute the the gradient potential terms among the remaining $v_i$ potential
terms then allows one to fine-tune the propagator to minimize the energy\cite{sak09,rota10}.

\begin{table}
\caption{Comparison of $N$ {\it spin-polarized} electron ground state energies 
$E_0/\hbar\w$ at coupling $\lambda=\sqrt{3}$ for $N=2$ (with exact
energy $E_0/\hbar\w=4)$ and at $\lambda=8$ for $N>2$.}
\begin{center}
\begin{tabular}{c r r r r r r r r}
\colrule
N &\ {\rm 8 PA-beads}&\ \ {\rm 2 beads}&\ {\rm 3 B-beads }&\  {\rm 4 B-beads } &\  {\rm 5 B-beads}& PIMC\cite{egg99} & CI\cite{ron06}& DMC\cite{ped03,gho07} \\
\colrule
2 &  4.042(5) &4.126(3)  &  4.033(2)      & 4.014(3)   & 4.001(4)   & & &         \\
3 & 15.694(3) &15.961(5)& 15.66(3)       & 15.63(3)   & 15.610(4)  & 15.59(1)  & 15.595 &         \\
4 & 27.92(1)  &28.266(5)& 27.898(4)      & 27.861(8)  & 27.82(2)   &\ \ \  27.823(11)& \ \ \ 27.828 &         \\
5 & 43.08(1)  &43.611(7)& 43.020(5)      & 43.00(3)   & 42.90(2)   & 42.86(4)  & 42.88 &         \\
\colrule
6 & 60.725(15)&61.403(7)& 60.622(6)      & 60.53(3)   & 60.46(2)   & 60.42(2)  & 60.80 & \ \ \ 60.3924(2)        \\
7 & 80.81(2)  &81.67(2)& 80.714(8)      & 80.59(2)   & 80.54(3)   & 80.59(4)  & & 80.5146(2)        \\
8 & 103.53(3) &104.45(1)& 103.42(1)      & 103.28(2)  & 103.18(3)  & 103.26(5) & & 103.0464(4)      \\
9 & 128.6(1)  &129.53(1)& 128.37(1)      & 128.23(4)  & 128.0(1)   & & &        \\
10 & 155.5(4) &156.77(1)& 155.38(1)      & 155.21(6)  & 154.9(2)   & & &         \\
\colrule
12 &        &217.55(2)& 215.79(2)      & 215.4(1)   & 215.2(2)   & & &         \\
14 &        &286.43(2) & 284.43(2)      & 284.08(8)  & 283.6(4)   & & &         \\
16 &          &363.08(2)& 360.53(5)      & 360.0(3)   & 359.5(6)   & & &         \\
18 &          &447.02(2)& 444.04(4)      & 442.9(4)   &            & & &         \\
20 &          &538.07(3)& 534.63(4)      & 534.1(2)   &            & & &         \\
\colrule
22 &          &635.99(2)& 632.06(4)      &    &            & & &         \\
25 &          &794.9(1)& 790.3(2)      &    &            & & &         \\
30 &          &1091.7(1)&      &    &            & & &         \\
\colrule
\end{tabular}
\end{center}
\label{tab1}
\end{table}

In Fig.\ref{compare} we
compare the Hamiltonian energy obtained by these optimized fourth-order propagator
with those of the PA propagator in Fig.\ref{pa}. The energy is computed by sampling 
the trace of only a {\it single} fourth-order propagator having 2 to 5 beads.
The 2-bead case, even without optimization, is substantially better than PA2. 
The optimized 3-bead case has energy lower than 
that of 8 PA propagators. The 5-bead case has energy lower than that of
Egger {\it et al}.'s calculation\cite{egg99}. As the calculation becomes more 
accurate with increasing number of beads, $E(\tau)$ levels off in approaching the
exact result, as in Fig.{\ref{ho}, but still retains a shallow minimum. 
Table \ref{tab1} compares our results to those obtained by the PIMC
method of Egger {\it et al.}\cite{egg99}, the configuration-interaction method
of Rontani {\it et al.}\cite{ron06} and the Diffusion Monte Carlo (DMC) method of
F. Pederiva	{\it et al.}\cite{ped03} and Ghosal {\it et al.}\cite{gho07}, 
for up to $N\approx 20$ spin-polarized electrons. 
In the case of 2 and 3 beads, results can be obtained up to $N=30$
and $N=25$ respectively. These calculations were 
done on a laptop computer and each entry can take up to $\approx 2-3$ days. 
Entries requiring longer running time were left undone. 
Substantial improvements are expected when the code is ported supercomputers.
In Fig.\ref{n20opt}, we show the convergence of the
3 and 4-bead propagators for solving the case of $N=20$ spin-polarized
electrons. This is a good illustration of the sudden appearance of the
sign problem, which blew up the 4-bead calculation with a large variance
at $\tau=7$. Nevertheless, the Hamiltonian estimator still gives
excellent upper-bounds to the energy at $\tau=6$ and $\tau=8$.

In this work, we have shown that optimized fourth-order propagators,
in using only 3-5 FFP, together with the use of the Hamiltonian estimator, 
can effectively limit the severity of the sign problem and allow accurate
calculation of quantum dot energies for up 20 fermions. The 2-bead calculations
are about 1\% too high, but they are without the sign-problem and can be used
as a quick variational estimate at $N$ much larger than 20.

I thank my colleague Eckhard Krotscheck for hospitality and support at 
Linz, Austria during the summers of 2010-2012, when this work was initiated.
This publication was made possible by NPRP GRANT \#5-674-1-114 
from the Qatar National Research Fund (a member of Qatar Foundation). 


\end{document}